\documentclass[runningheads]{svmult}

\usepackage{makeidx}   
\usepackage{graphicx}  
\usepackage{subeqnar}  
\usepackage{multicol}  
\usepackage{physprbb}  
\makeindex             

\def\gtsima{$\; \buildrel > \over \sim \;$}
\def\ltsima{$\; \buildrel < \over \sim \;$}
\def\prosima{$\; \buildrel \propto \over \sim \;$}
\def\gsim{\lower.5ex\hbox{\gtsima}}
\def\lsim{\lower.5ex\hbox{\ltsima}}
\def\simgt{\lower.5ex\hbox{\gtsima}}
\def\simlt{\lower.5ex\hbox{\ltsima}}
\def\simpr{\lower.5ex\hbox{\prosima}}

\def\h1{$h^{-1}$}
\def\eeq{\end{equation}}
\def\beq{\begin{equation}}

\begin{document}
\title*{Tracing the formation of massive spheroids\protect\newline 
from high-$z$ galaxy clustering}
\toctitle{Tracing the formation of massive spheroids
\protect\newlinefrom high-$z$ galaxy clustering}
%
%
\titlerunning{High-$z$ clustering and spheroid formation}
%
\author{Emanuele Daddi\inst{1}
%
\authorrunning{Emanuele Daddi}
%
%
\institute{European Southern Observatory, Karl-Schwarzschild-Str.~2,
D-85748 Garching bei M\"unchen}
}

\maketitle              

\begin{abstract}
The high-$z$ progenitors of local massive early-type galaxies should
be characterized by a strong level of clustering, larger than that
measured for $z=3$ Lyman Break Galaxies and comparable to that of $z\sim1$
EROs. First possible evidences for such strongly clustered objects
at $z\simgt2$ were found by the FIRES and K20 surveys, that have identified
new classes of faint high-$z$ K-selected galaxies. Some details are given
here for the new population of massive star-forming galaxies at
$z\sim2$, found by the K20 survey in the GOODS-South area. 
Because of their much redder UV continuum, most of these galaxies would 
not be selected by the Lyman Break criterion.
Such objects are good candidates for
the precursors of local ellipticals caught in their formation phase.
We have calibrated a two color criterion to allow the identification
of these highest redshift galaxies in bright K-limited samples.
\end{abstract}

\section{Introduction}
Understanding the evolution, star-formation and assembly histories
of early type galaxies up to their highest redshifts of formation
is a crucial observational question that remains unsolved.
Passively evolving, massive spheroidal galaxies appear to be common up
to $z\approx 1$ with little evolution, if any, from present days
\cite{D00,pozz,bell}. It is not well constrained yet if
and how many passive spheroids exist at higher redshifts, $z>1.5$--2.
Moreover, it is unclear up to what redshift most precursors of 
spheroids maintain the low-redshift signatures of morphological 
and stellar population properties. 
Approaching their formation epoch, spheroid progenitors 
will
appear as star-forming and possibly morphologically irregular galaxies.
Therefore, identifying the precursors of todays ellipticals close to their
most interesting formation or assembly phases requires to rely on independent 
and alternative signatures than their morphologies or old stellar
population properties.

\section{Expected clustering of spheroid precursors}

\begin{figure}[ht]
\begin{center}
\includegraphics[width=.8\textwidth,angle=0]{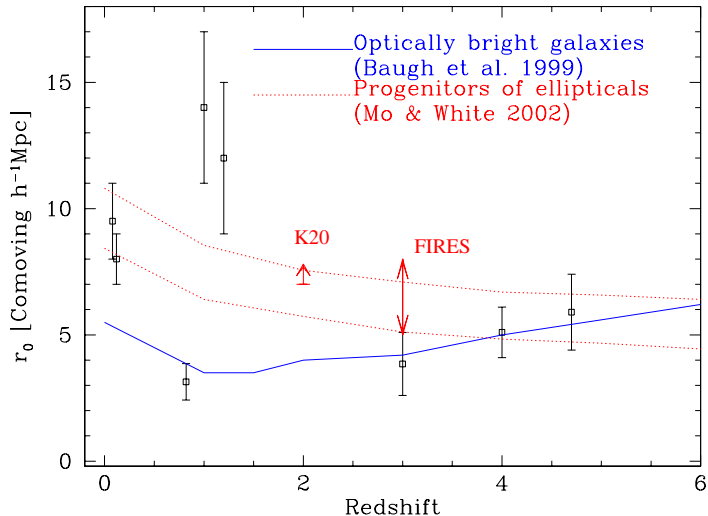}
\vspace{-0.5truecm}
\end{center}
\caption[]{Predicted evolution for the correlation length of halos
that will evolve into local massive ellipticals (dotted lines, from
\cite{mowhite}) and of optically selected galaxies (solid line,
from \cite{baugh99}). 
Together with the new constraints coming 
from K-selected populations at $z\simgt2$ (heavy symbols), we 
show relevant measurements taken from the literature: ellipticals 
at $z\sim0$ from \cite{norberg}, optically selected galaxies 
at $z\sim0.8$ from \cite{coil}, EROs and radio galaxies at $z\sim1$ 
from \cite{D01,Ov},
$z=3$ LBGs from \cite{GD01} and $z=4$--5 LBGs from \cite{ouchi}
}
\label{fig1}
\end{figure}

Spheroids inhabit preferentially the densest environment in the
local universe.
They have the largest correlation lengths among galaxies in the
present day universe and up to $z\sim1$ \cite{norberg,D01}. 
{\em Searching for the most clustered population at any
redshift is therefore a natural way to locate such progenitors
and to investigate their properties.}

This view is supported by model predictions for the redshift
evolution of the correlation length of halos  
hosting today massive ellipticals (Fig.~\ref{fig1}, based on \cite{mowhite}).
Halos clustered with $r_0\sim8$--10 in the local universe, are 
predicted to have $r_0\sim 5$--8 \h1 Mpc even at $2<z<4$.
Being referred to halos, these predictions  could represent
only lower limits for galaxies that, at the relatively small
scales probed by high-$z$ measurements, can have enhanced correlations
due to large halo occupation numbers (e.g. \cite{M&S}).

In practice, it is impossible to observationally preselect a population 
by its clustering properties and one has to look at the clustering
properties of the available classes of high-$z$ galaxies to find
putative progenitors.

Extremely red objects (EROs) and giant $z\sim1$ radio galaxies, as well 
known, have an high clustering and therefore are good candidate
progenitors of local ellipticals \cite{D00,McC,D01,Ov}, 
in agreement with some
significant fraction of these objects being 
passive spheroids (e.g. \cite{C03}).

Measurements for the
highest redshifts ($z\sim4$--5) star-forming
galaxies \cite{ouchi} of $r_0\sim5$--6 may be consistent with the 
predictions for 
forming spheroids. This is also in qualitative agreement with the
observation that the oldest stars (thus the ones formed at the highest
redshifts) are nowadays in spheroids.

Typical $z=3$ Lyman Break Galaxies (LBGs) appear instead to start 
falling short
of the expected clustering level of ellipticals progenitors, with the
best current estimates of $r_0\simlt4$ \h1 Mpc \cite{steidel} and
with fainter LBGs having even much weaker clustering \cite{GD01}.
In fact the clustering of observed LBGs at $z=3$ is much lower than that of 
$z\sim1$ EROs, suggesting that possibly 
a large fraction of them do not evolve into EROs and ellipticals.
Overall, the observed clustering of optically selected 
(i.e. actively star forming)
galaxies from $z=5$ to $z=1$ \cite{coil} is in very good agreement with
the model predictions of \cite{baugh99}. The comparison of the two set
of predictions of Fig.~\ref{fig1}, diverging below $z\sim4$,  
also confirm that by $z\simlt 3$
ordinary star formation has began producing stars that have mainly ended
in today later type or less massive galaxies.

Should it be possible to isolate the progenitors of local spheroids
as a class of galaxies at $1.5<z<3.5$, their clustering would be expected
to be fairly strong, in the cited range $r_0\sim5$--8 \h1 Mpc or even more.
New populations of K-selected, $z\sim2$--3 galaxies 
possibly satisfying those clustering requirements were recently reported 
by the FIRES and K20 surveys \cite{D03,D04}. 
Along with SCUBA sources, these galaxy populations represent the best 
candidates so far identified for the precursors of massive spheroids.

\section{$J-K$ red FIRES galaxies at $2<z<4$}

The FIRES survey of the HDF-South \cite{labbe03} has produced 
the first large sample of about 120 $K$-selected galaxies in the 
(photometric) redshift range $2<z<4$. 
From the analysis of their correlation properties \cite{D03}, these
$z=3$ K-selected galaxies appear more strongly clustered than LBGs 
(at a fixed number density), although the comparison suffers from
the small area of HDFS. 
A significant result was the detection in the FIRES sample
of a strong color dependence of clustering, in the usual way
that redder galaxies appear more clustered (but a previously unobserved effect
at any $z>1$). The clustering depends primarily on the
$J-K$ color, which brackets the Balmer (or
4000\AA) break at $z\approx 3$, suggesting the effect could be due to
age. Intriguingly, the existence of populations older and more clustered
than LBGs could solve the paradox that $z=3$ LBGs are too young to be 
direct progeny of $z>4$--6 star forming galaxies \cite{ferg02}.

The galaxies with the reddest ($J-K>1.7$) colors at $z\approx 3$ have
an angular two-point correlation function consistent with $r_0\sim 8$ 
\h1 Mpc (using a typical slope $\delta=0.8$), although a weaker $r_0\simgt5$
\h1 Mpc could be made consistent with the data assuming enhanced small scale
correlations in the "halo occupation distribution" framework 
\cite{zheng,Kravtsov}. 
Overall, these clustering strengths of 5--8 \h1
Mpc are at the level expected for the progenitors of ellipticals seen
at $z\sim3$ (see Fig.~\ref{fig1}). 
The first handful spectroscopic redshifts for this class of $J-K$ red
galaxies \cite{pvd} are grouped in a single redshift spike,
a feature that support the presence of strong clustering.

\section{Near-IR bright galaxies at $z\sim2$ from the K20 survey}

\begin{figure}[t]
\begin{center}
\includegraphics[width=.8\textwidth,angle=0]{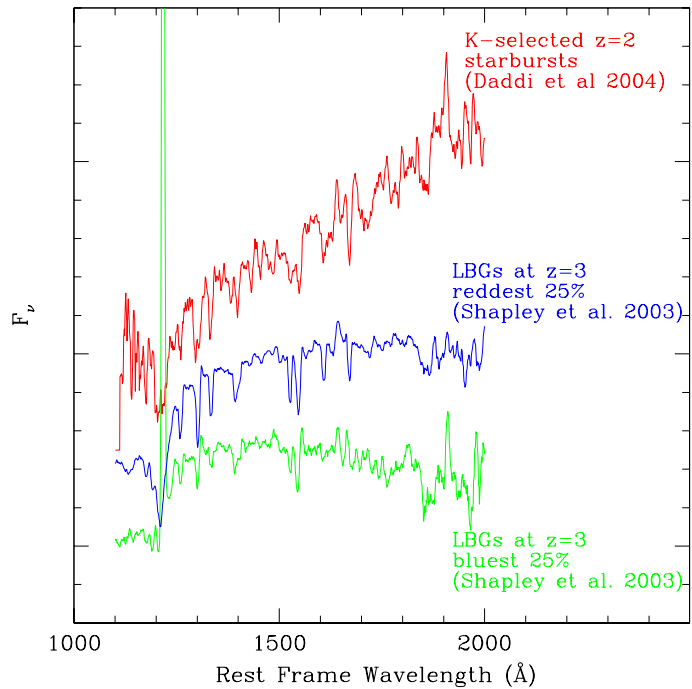}
\vspace{-0.5truecm}
\end{center}
\caption[]{The coadded spectrum of the $z=2$ K-selected galaxies
(top, from \cite{D04}) is compared to the reddest and bluest composite
spectrum of LBGs (center and bottom, from \cite{shap03}). The strong 
Ly$\alpha$ emission is from the bluest LBG spectrum.
}
\label{fig3}
\end{figure}

The K20 survey resulted in the observation of a significant high redshift
tail of $1.7<z<2.3$ galaxies with bright near-IR magnitudes ($K_{Vega}<20$,
\cite{C02}).
The vast majority of such galaxies appear to be actively star-forming,
reddened, morphologically irregular and massive objects \cite{D04}.

Remarkably, 6 out of the 9 galaxies with measured spectroscopic redshift belong
to spikes with $\simlt1500$ km/s of redshift separation. 
Assuming a random redshift selection function within the $1.7<z<2.3$ range
such a level of pairing has a probability of about $10^{-5}$,
a clear evidence of clustering.
It is intriguing that even for EROs a similarly strong level of 
redshift pairing is not observed.
We performed Monte Carlo simulations building samples of clustered populations
(using the method presented in \cite{D02}) in order to see what 
correlation length is necessary to produce a similar redshift pairing
in our small sample.
It is found that in order to have a 5\% or more probability to find 
as many pairs
as in the observed sample, $r_0\simgt7$ \h1 Mpc is required. 
The fraction of close pairs in a survey is directly related to
the integral of the real space two-point correlation function, and has little 
dependence on the correlation slope $\delta$. The pairing evidence thus
suggest that these near-IR bright $z\sim2$ galaxies are consistent
with having a high-level of clustering, like the one expected for the 
progenitors of local ellipticals seen at $z\sim2$. 
It would be extremely interesting to spectroscopically confirm a much larger
samples, in order to obtain improved estimates for the clustering
of this galaxy population.

\begin{figure}[t]
\begin{center}
\includegraphics[width=.8\textwidth,angle=0]{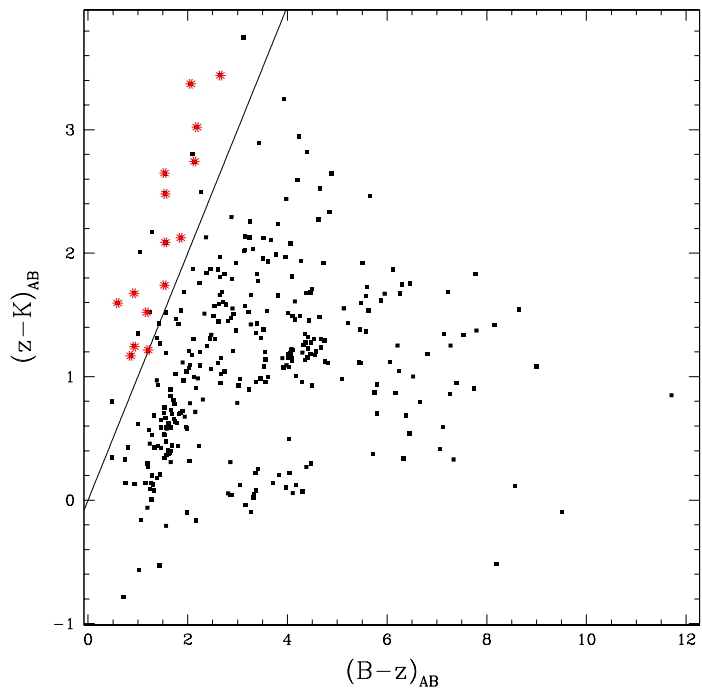}
\vspace{-0.5truecm}
\end{center}
\caption[]{Two color diagrams showing galaxies from the K20 survey in the
CDFS field with the 9 near-IR bright galaxies at $z\sim2$ shown as asterisks,
including also a few redder objects with reliable photometric redshift $z>2$.
The straight line has $(z-K)_{AB}=(B-z)_{AB}$. 
}
\label{fig4}
\end{figure}

Galaxies at $z\sim2$ with bright K-band magnitudes
were unknown before \cite{D04} and \cite{pvd}, and actually most theoretical
models had predicted that they should not exist at all (see e.g. \cite{C02} for
relative references). As they appear to be actively star-forming, just like
LBGs, it is interesting to assess whether these K-selected galaxies are just
a special subsample (the brightest tip of the iceberg)
of LBGs, or a different population. This is especially 
important as now large samples of $\sim1000$ LBGs have been assembled
down to $1.5<z<2.5$ \cite{steidel}.
Fig.~\ref{fig3} shows that the coadded spectrum of $z=2$ K-selected galaxies is
remarkably redder than typical LBGs at $z=3$, even for the reddest among them.
For a more direct comparison to LBGs, we used 
the available multicolor photometry in the CDFS/GOODS area to determine the
synthetic {\em G} and {\em R} magnitudes of our 9 $z=2$ objects, 
matching the $z=2$ selection criterion 
of \cite{steidel} (the available U-band
data closely match the system of \cite{steidel}). Although the procedure
is a bit uncertain, the {\em G}$-${\em R} colors of near-IR bright $z\sim2$ 
objects appear again generally much redder than what required by the 
LBG selection criterion in the same redshift range \cite{erb}. 

These near-IR bright $z=2$ galaxies are thus a new population
of high-$z$ galaxies, that cannot be selected with the usual LBG criterion.
As they are only $\approx5$\% of the population in a K-limited sample
at $K<20$, it would be very useful to have an alternative method to
distinguish them from lower redshift field galaxies. Using the highly complete
spectroscopic sample of the K20 survey we have calibrated a very efficient 
two color criterion: galaxies with $(z-K)_{AB}\simgt(B-z)_{AB}$ have an
high probability to be star-forming objects at $z\sim1.5$--2.5 
(Fig.~\ref{fig4}).
We are currently using this {\em BzK} criterion to build up a much
larger sample of spectroscopically identified near-IR bright $z=2$ galaxies.
This will be crucial in order to understand their
properties and the role played in the formation history of galaxies
and to confirm their nature of possible star forming progenitors of
local spheroids.

\smallskip
\noindent
{\bf Acknowledgments.}
{\footnotesize
The K20 and FIRES teams are warmly thanked. The work
presented here would have not been possible without the dedicated
efforts of all the people involved in those projects.
}

%

\end{document}